\documentclass[superscriptaddress, preprint]{revtex4-2}
\usepackage{graphicx}
\graphicspath{ {../fig/}}

\usepackage{amssymb}
\usepackage{amsmath}
\usepackage{amsthm}
\usepackage{amsfonts}
\usepackage{xfrac}
\usepackage{wasysym}
\usepackage{textcomp}
\usepackage{bbm}

\usepackage{braket}
\usepackage{mhchem}
\usepackage{epstopdf} 

\usepackage{hyperref}
\usepackage{array}
\begin{document}

\author{Stefano Marin}
\email{stmarin@umich.edu}
\affiliation{Department of Nuclear Engineering and Radiological Sciences, University of Michigan, Ann Arbor, MI 48109, USA}


\author{M. Stephan Okar}
\affiliation{Department of Nuclear Engineering and Radiological Sciences, University of Michigan, Ann Arbor, MI 48109, USA}





\author{Shaun D. Clarke}
\affiliation{Department of Nuclear Engineering and Radiological Sciences, University of Michigan, Ann Arbor, MI 48109, USA}


\author{Sara A. Pozzi}
\affiliation{Department of Nuclear Engineering and Radiological Sciences, University of Michigan, Ann Arbor, MI 48109, USA}
\affiliation{Department of Physics, University of Michigan, Ann Arbor, MI 48109, USA}

\title{Generalization of the Maier-Leibniz Doppler-Shift Method for Gamma-Ray Correlations in Fission}
\date{\today}

\begin{abstract}
The Maier-Leibniz Doppler-Shift technique is the most popular and accurate technique used in fission experiments to separate the yield of gamma rays from each of the two fission fragments. The technique exploits the aberration, i.e., the change in the angular distribution, of gamma rays emitted by a moving source. By measuring the speed and direction of the source with a conventional detector, as well as the yield of gamma rays at several angles from the direction of motion, the technique can be used to determine the mean multiplicities of gamma rays from each fragment. We show in this work that it is possible to extend the technique to also measure second moments of the gamma ray radiation from each fragment. In particular, given the current interest in fragment correlations in fission, we show that the covariance of the emission between the two fragments can be inferred. Experimental limitations and convergence of the new technique are discussed. 
\end{abstract}


\maketitle

\section{Introduction} 
\label{sec:intro}

A characteristic feature of the binary fission process is the presence of two excited nuclear fragments moving antiparallel to one another at moderate velocities, $\approx 0.05 \ c$. The two fragments quickly de-excite by emission of neutrons, followed by gamma rays in the continuum, and finally gamma rays in the discrete-level region until the ground state, or a long lived isomeric state, is reached. Over the past few years, a renewed interest in the fission reaction was sparked by new measurements and refined theoretical models~\cite{Travar2021, Wilson2021, Schmidt2011, Talou2018}. Of particular interest are the event-by-event correlations between the fission fragment properties. 

Neutron emission is relatively easy to separate between the two fragments. The evaporated neutrons have velocities comparable in magnitude to the speed of the fragments, resulting in strong kinematic focusing of the emission along the fission axis. By analysing both energy and direction of the neutrons and the fragments, it is possible to determine with decent accuracy which fragment emitted each neutron on an event-by-event case~\cite{Signarbieux1972, nifeneckergroups, Gavron1971}. Similarly, the discrimination of discrete gamma rays can be performed by gating on specific known transitions of nuclei~\cite{Wilson2021}. The same is not possible in general for statisical gamma-rays in the continuum. However, as shown by Maier-Leibniz~\cite{MaierLeibniz1965} and Pleasonton \textit{et al.}~\cite{Pleasonton1972}, the gamma-ray emission can be analysed using statistical methods such as Maier-Leibniz Doppler-Shift (ML-DS) technique. 

The (ML-DS) technique has been successfully applied successfully in several experimental investigations of fission, including Refs.~\cite{Pleasonton1972, SchmidFabian1988, Wang2016, Travar2021}. However, given the recent interest in fragment correlations (see Refs~\cite{Wilson2021, Vogt2021, Bulgac2020, Schmidt2011}) we think it is important to present techniques capable of determining correlations between gamma-ray emission from the two fragments. In this paper, we will show that a simple generalization to the ML-DS technique allows the experimenter to separate the second order moments of the gamma-ray distribution without the need of changing the experimental apparatus. 

\section{Overview of Technique}
\label{sec:overview}

The ML-DS technique has been applied in several experiments and several variants of the techniques exist. In its original formulation~\cite{MaierLeibniz1965, Pleasonton1972}, the method only employs a single gamma ray detector. One of the main recent variants of this technique, from Travar \textit{et al.}~\cite{Travar2021}, used two detectors in order to eliminate the effect of a fragment detector asymmetry. This version is briefly discussed here.

Let $N^L$ and $N^H$ be the random variables describing the gamma-ray multiplicities from the light and heavy fragments, respectively. Let us then take two detectors placed along the line of motion of the fragments, such that the light fragment flies in the direction of detector $I$ and the heavy fragment flies in the direction of detector $II$. Let $D^I$ and $D^{II}$ be the random variables describing the measured gamma-ray multiplicity distributions in detectors $I$ and $II$, respectively. Assuming that each gamma ray is independently detected , and thus the system response can be modeled as a binomial response, we have

\begin{subequations}
    \begin{align}
    D^I &=& \hat{B}(\epsilon_{L+})N^L + \hat{B}(\epsilon_{H-})N^H \\
    D^{II} &=& \hat{B}(\epsilon_{L-})N^L + \hat{B}(\epsilon_{H+})N^H 
    \end{align}
\label{eq:randD}
\end{subequations}
where $\hat{B}(\epsilon)$ is the binomial response operator that models the effect of a system efficiency in measuring multiplicities, and the Doppler-corrected detection efficiency

\begin{equation*}
     \epsilon_{F\pm} = \epsilon (1 \pm 2 \beta_F) \ ,
\end{equation*}
is given in terms of the detection efficiency $\epsilon$ of each detector to an isotropic source and the speed $\beta_F$, as a ratio of the speed of light, of the fragment $F = L,H$ emitting the radiation. Taking the mean of Eq.~\eqref{eq:randD}, we find the well-known formula for inferring the mean gamma-ray emission~\cite{Travar2021}

\begin{subequations}
\begin{align}
    \langle N^L \rangle & = & \frac{ (1+ 2 \beta_H)\langle D^{I} \rangle - (1 - 2 \beta_H) \langle D^{II} \rangle}{4 \epsilon (\beta_L + \beta_H)}  \\ 
    \langle N^H \rangle & = & \frac{ (1 + 2 \beta_L)\langle D^{II} \rangle - (1 - 2 \beta_L) \langle D^{I} \rangle}{4 \epsilon (\beta_L + \beta_H)}  \ . 
\end{align}
\label{eq:meanEmit}
\end{subequations}
Eq.~\eqref{eq:meanEmit} is the conventional ML-DS technique. We now extend its use by obtaining the covariance between the emitting sources. Analysing the second order moments of Eq.~\eqref{eq:randD} distributions, we find that they are related to the emitted multiplicity distribution by the following 

\begin{equation}
    \begin{bmatrix}
\Sigma^2(D^I) \\
\Sigma^2(D^{II}) \\
\text{cov}(D^I, D^{II})
\end{bmatrix}
= 
\begin{bmatrix}
\epsilon_{L+}^2 & \epsilon_{H-}^2 & 2 \epsilon_{L+} \epsilon_{H-} \\
\epsilon_{L-}^2 & \epsilon_{H+}^2 & 2 \epsilon_{L-} \epsilon_{H+} \\
\epsilon_{L+} \epsilon_{L-} & \epsilon_{H-} \epsilon_{H+}  & \epsilon_{L+}\epsilon_{H+} + \epsilon_{L-}\epsilon_{H-}
\end{bmatrix}
\begin{bmatrix}
\sigma^2(N^L) \\
\sigma^2(N^H) \\
\text{cov}(N^L,N^L) 
\end{bmatrix} \ ,
\label{eq:matLin}
\end{equation}
where we have introduced the reduced variances,

\begin{subequations}
    \begin{align}
    \Sigma^2(D^I) &=& \sigma^2(D^I) - \langle N^L \rangle \epsilon_{L+} (1 -\epsilon_{L+}) -\langle N^H \rangle \epsilon_{H-} (1 -\epsilon_{H-}) \\ 
   \Sigma^2(D^{II}) &=& \sigma^2(D^{II}) - \langle N^L \rangle \epsilon_{L-} (1 -\epsilon_{L-}) - \langle N^H \rangle \epsilon_{H+} (1 -\epsilon_{H+}) \ .
    \end{align}
\end{subequations}
The reduced variance has a physical meaning. We subtract the expected variance introduced by the binomial operator from the variance observed in the detectors. This subtraction can be understood in terms of removing the noise associated to an information channel, i.e., the binomial response. The mean emitted multiplicities appearing in the reduced variances can be determined using Eq.~\eqref{eq:meanEmit}. 

After inverting the matrix equation in Eq.~\eqref{eq:matLin}, we obtain an expression for the covariance in the emission from the two fragments
\begin{equation}
    \text{cov}(N^L, N^H) = \frac{2 \text{cov} \left(D^I, D^{II} \right) (1 + 4 \beta_H \beta_L ) - \left[ (1 + 2 \beta_H ) (1 - 2 \beta_L) \Sigma^2(D^I) +(1- 2 \beta_H ) (1 + 2 \beta_L) \Sigma^2(D^{II}) \right]}{16 \epsilon ^2 (\beta_L + \beta_H)^2} \ .
\label{eq:varD}
\end{equation}

We note that Eq.~\eqref{eq:varD} represents the difference of two factors similar in size, which is then amplified by the division of the very small factor in the denominator. For this reason, the equation has a very slow convergence and requires high statistics. Numerical analysis using reasonable values for the emission, efficiencies, and speed of the fragments, only converge after approximately $1\times 10^8$ events, thus requiring extremely long measurement times.

A factor that affects the determination of the covariance, and to a lesser extent the mean, is the capability of the detector of measuring the multiplicity. If only detectors capable of measuring a single interaction per event are used, both the mean and the covariance expression require that the probability of two incident particles on the detector be much smaller than 1. This restriction further increases the measurement time. To increase the efficiency of the system, it is possible to place detectors off of the fragment-motion axis. However, the efficiency of each detector would have to be modeled independently because the aberration of gamma-rays, apart for points along the fission axis, is in general dependent on the angular distribution of gamma rays in the inertial frame of the fragment.

Notwithstanding these limitations, as far as we know Eq.~\eqref{eq:varD} represents the only technique capable of measuring the covariance between the emission of gamma rays by two sources moving at relatively small speeds. This technique represents the missing piece in the experimental analysis of fission emission correlations. Together with the techniques discussed for separating the emission of neutrons and discrete gamma rays, this technique will allow to determine the correlations in the initial conditions of the fission fragments and gain a deeper understanding of the fission process. 

\acknowledgments

This work was funded in-part by the Consortium for Monitoring, Technology, and Verification under Department of Energy National Nuclear Security Administration award number DE-NA0003920. 

\bibliography{mybib}



\end{document}